\documentclass[openacc]{rstransa}%%%%where rstrans is the template name

\usepackage{color}

\def\DELME#1{{\textcolor{green}{}}}         % deletion 
\def\DEL#1{{\textcolor{green}{ }}}          % suggested deletion in text
\titlehead{Research}

\begin{document}

%%%% Article title to be placed here
\title{Dynamical and statistical properties of large scales in turbulence}

\author{%%%% Author details
A. Alexakis$^{1}$, M. E. Brachet$^{1}$,  S. Fauve$^{1}$ and F. P\'etr\'elis$^{1}$}

%%%%%%%%% Insert author address here
\address{$^{1}$Laboratoire de Physique de l'Ecole Normale Sup\'erieure, CNRS, PSL Research University,
Sorbonne Universit\'e, Universit\'e Paris Cit\'e, F-75005 Paris, France}
%$^{2}$Second author address\\
%$^{3}$Third author address}

%%%% Subject entries to be placed here %%%%
\subject{fluid dynamics, statistical physics}

%%%% Keyword entries to be placed here %%%%
\keywords{turbulence, equipartition, transitions between turbulent regimes}

%%%% Insert corresponding author and its email address}
\corres{Stephan Fauve\\
\email{stephan.fauve@phys.ens.fr}}

%%%% Abstract text to be placed here %%%%%%%%%%%%
\begin{abstract}

We present a review of results concerning the properties of scales larger than the forcing scale in turbulent flows. In three-dimensional turbulence, when the power driving the flow is injected at a well-defined scale, the injection scale separates the small-scale range, where the Kolmogorov cascade takes place, from the large-scale range across which the mean energy flux is zero, suggesting that these modes are in equilibrium. We show that, in the case of spatially periodic forcing involving only a few modes, an equipartition energy spectrum is generated at large scales, and we discuss the deviations from equipartition observed with more complex forcings.
In two-dimensional turbulence, the injection scale separates the direct enstrophy cascade toward small scales from the inverse energy cascade toward large scales. Because of the inverse cascade, one would not expect equilibrium statistical physics tools to be suitable for describing the large-scale structures. Yet this is precisely the situation in which such tools have been most widely used. We indeed show that the bifurcations observed between different large-scale regimes of two-dimensional turbulence, both in experiments and in direct numerical simulations, can be qualitatively described by the Truncated Euler equations, whose solutions satisfy statistical equilibrium.

\end{abstract}
%%%%%%%%%%%%%%%%%%%%%%%%%%%

%%%%%%%%%% Insert the texts from introduction below which can accomdate on firstpage in the tag "fmtext" %%%%%

\begin{fmtext}
%\NOTE{You can put the start of the intro (including the section latex command), just make sure that it will fit here on the first page...}
\end{fmtext}
%%%%%%%%%%%%%%% End of first page %%%%%%%%%%%%%%%%%%%%%

%%%%%%%%%%%%%%% End of first page %%%%%%%%%%%%%%%%%%%%%
\maketitle

\section{Introduction}

Tools from statistical mechanics and probability theory have been applied for a long time to the analysis of a turbulent signal. A turbulent velocity field indeed possesses many different scales and is unpredictable. This means that the details of the velocity field are continuously changing in an incoherent way. However, mean values or more generally probability density functions are constant when the flow is submitted to a statistically stationary forcing. The methods of stationary random processes are therefore appropriate and have been first used in the context of turbulence by G. I. Taylor \cite{taylor1935} and are reported in detail in well-known books \cite{batchelor1953, monin1975}.

 The equation governing the probability density function of the velocity field for three-dimensional turbulence, has been derived by Hopf who tried to solve it in particular cases. He emphasized that the solution strongly depends on the existence of viscous dissipation. In the absence of dissipation, he showed that the velocity field is Gaussian and the velocity modes are in equipartition, thus leading to a spectrum proportional to $k^2$ where $k$ is the wave number \cite{Hopf1952}. This result was also obtained by Lee using the truncated Euler equations  \cite{lee1952}. Later, Kraichnan took into account the effect of the flow helicity which is a second conserved quantity by the three-dimensional truncated Euler equations besides the energy \cite{kraichnan1973}. 
 
 The truncated Euler equations are a set of coupled ordinary differential equations for the amplitudes of the velocity modes with wave vectors contained inside a sphere of finite radius $k_{\rm max}$. This dynamical system obeys a Liouville equation.  Therefore, assuming that all configurations in phase space are equally probable, subject only to the constraints of energy and helicity conservation, one obtains the microcanonical distribution at equilibrium. When no single mode carries a significant fraction of the total energy and a sufficiently large number of modes are excited, the canonical or the grand canonical ensembles can then be used. It is within this framework that Kraichnan computed the energy and helicity spectra. \cite{kraichnan1973}.
 
 The same approach can be used for two-dimensional flows governed by the truncated Euler equations. The conserved quantities are then the kinetic energy and the enstrophy (the mean square vorticity). The system displays richer structure than for the three-dimensional case \cite{kraichnan1980}. There exist equilibrium solutions that involve negative temperatures  in which the energy is mostly concentrated into the largest scales of the flow. Equilibrium statistical mechanics of the two-dimensional Euler equations has also been used in real space, first by Onsager, who considered flows that consist of discrete vortices \cite{onsager1949}. The generation of large scales resulting from the merging of equal sign vortices in a negative temperature state was also observed. This approach was extended to continuous solutions of the two-dimensional Euler equations for which all conserved quantities have been taken into account \cite{miller1990,robert1991}. Therefore, statistical mechanics of the two-dimensional Euler equations seem to provide a possible explanation for the generation of large scale structures observed in two-dimensional turbulence. This result is a priori surprising because we do not expect the system to be at statistical equilibrium due to the presence of an inverse cascade of energy, phenomenologically described by Kraichnan \cite{kraichnan1967}. 
 %This inverse cascade is another possible explanation for the generation of large scales. This needs to be discussed.
 Indeed, when a large scale drag force is present with a drag coefficient sufficiently large to dissipate energy 
 before it reaches the domain-size scales, a Kolmogorov $k^{-5/3}$ spectrum is observed that marks the presence of the inverse cascade. If however the drag coefficient is too weak or absent, energy  piles up in the domain size scales leading to much steeper spectra. At such a state the fluctuations of the energy flux are much larger than their mean (negative) value and one can argue that the flows are in quasi-equilibrium with the mean flux playing a secondary role \cite{vankan2025}.
 
It has been emphasized in the literature that statistical equilibrium solutions obtained for the Euler equations (zero viscosity) strongly differ from the observations of real Navier-Stokes turbulence \cite{Hopf1952}. All the absolute equilibrium spectra (no viscosity) differ indeed from the spectra observed in two-dimensional or three-dimensional turbulence. In their 1989 review paper, ''Is there a statistical mechanics of turbulence?"\cite{kraichnan1989}, Kraichnan and Chen conclude that although approximations such as the Direct Interaction Approximation (DIA) provides a good description of the truncated Euler equations, it becomes less adequate when viscosity is taken into account. It could approximately predict spectra but it fails to capture intermittency. The status is similar for renormalization group methods that have a long story in the field of turbulence \cite{forster1977,fournier1978,dedominicis1979, yakhot1986,canet2016}. Although recent progress has been made, some results still lack rigor and intermittency of three-dimensional turbulence cannot be described. 

 The existence of power-laws and in particular of non classical scaling related to intermittency have motivated analogies between fully developed turbulence and phase transitions with anomalous exponents \cite{nelkin1974,nelkin1975,degennes1975}. Although these analogies were perhaps useful to motivate the use of the renormalization group methods, they are limited and have not provided tools to explain or predict basic features of turbulent flows. \\

Finally, Kolmogorov-type power-laws, associated with energy or wave-action fluxes and equilibrium spectra, have been extensively studied in wave turbulence, that is, in systems involving a set of weakly nonlinear interacting waves \cite{nazarenko2011}. From this point of view, wave turbulence exhibits strong analogies with hydrodynamic turbulence.
Some systems, such as capillary waves for example, involve only a direct cascade of energy toward small scales and no inverse cascade, in a manner analogous to three-dimensional hydrodynamic turbulence. Others, such as gravity waves, exhibit an inverse cascade of wave action, which makes them more similar to the case of two-dimensional turbulence.\\

The above short review shows that many tools from statistical physics have been used to try to explain turbulence, with partial but limited success. A more modest scientific objective has been considered for some years now. For turbulent flows generated by a forcing with a well-defined spatial scale, it consists of asking whether spatial scales larger than the forcing scale can be described using statistical mechanics at equilibrium, even if these scales are coupled to those of the inertial zone, which is a system strongly out of equilibrium. We review here recent works performed in this direction.

\section{Statistical properties of large scales in three-dimensional turbulence or wave turbulence with a direct cascade}

Most experimental and numerical studies of three-dimensional turbulence have focused on the inertial and dissipative ranges of the flow, that is, on scales smaller than the forcing scale, where the Kolmogorov cascade transfers energy toward smaller scales in the inertial range before it is dissipated below the Kolmogorov scale. In contrast, spatial scales larger than the forcing scale have received far less attention.

From a technical standpoint, this oversight can be explained. In direct numerical simulations, forcing the flow at one of the lowest available modes in the computational domain maximizes the separation between the forcing and dissipative scales, allowing for the highest possible Reynolds number. In laboratory experiments, the size of the container is typically of the same order as the forcing scale, except in specially designed setups. 

Yet, many three-dimensional flows of geophysical or astrophysical relevance involve spatial structures much larger than the forcing scale.
It is also possible to design laboratory experiments in which the flow is driven by a spatially periodic force with a wavelength much smaller than the container size, known as Kolmogorov-type forcing, even in a three-dimensional configuration. Similarly, in technological applications, mixing often involves arrays of propellers, and wind turbine farms provide a recent example of flow structures extending beyond the size of individual turbines or the spacing between neighboring ones.

Studying the behavior of turbulence at scales larger than the forcing scale is therefore of both fundamental and practical interest. It is thus surprising that this problem has not received more attention in the turbulence literature. The topic is very briefly mentioned in Frisch's book on turbulence, where he notes that, although seemingly unphysical, absolute equilibrium solutions of the Euler equation \cite{Hopf1952,lee1952,kraichnan1973} might be “appropriate at the very smallest wavenumbers of turbulent flow maintained by forcing at intermediate wavenumbers,” as shown by Forster, Nelson and Stephen in \cite{forster1977}.

Indeed, that study demonstrated that when a Gaussian white-noise forcing is applied at an intermediate spatial scale in the Navier–Stokes equations (model C), the energy spectrum at low wavenumbers behaves as though the largest scales are in equipartition, yielding $E(k) \propto k^2$. It is important to clarify - contrary to certain incorrect claims in the literature - that Hopf, Lee, Kraichnan, and others who studied absolute equilibrium solutions of the Euler equation never considered them as possibly describing the behavior of large scales in turbulent flows of viscous fluids.

It is therefore of interest to investigate whether the results of \cite{forster1977} persist under more realistic types of forcing than Gaussian white noise. A plausible argument can be made: in three-dimensional turbulence forced at a scale $l_f$ at which viscous dissipation is negligible, within a domain of size $L \gg l_f$, the energy injected at $l_f$ cascades only toward smaller scales, leading to a zero mean energy flux in the range $l_f < l < L$. The larger-scale modes thus exchange energy with those in the inertial range $l < l_f$,  but without a net energy flux between them. As a result, these large-scale modes may reach a statistical equilibrium, potentially leading to equipartition and the corresponding spectrum $E(k) \propto k^2$.

It is worthwhile at this stage to clarify what is meant by equilibrium. A system is in thermal equilibrium if the probability laws that characterize it do not evolve over time and, moreover, if it is not crossed by any macroscopic flux with a nonzero mean value.
In the case of a viscous fluid, energy must be transferred from the scale at which it is injected to the scales at which dissipation is significant. There is therefore an energy flux that drives the system out of equilibrium. In the case of a hypothetical inviscid fluid, no nonzero mean flux is expected in the absence of forcing.
A more stringent definition of thermal equilibrium relies on the principle of detailed balance, which states that every microscopic transfer of energy is counterbalanced by the reverse transfer occurring with the same probability. This excludes flux-loop states in which the downscale and upscale energy fluxes are related to different mechanisms such that the mean flux vanishes when a steady state is reached \cite{boffetta2011,xie2019,clark2020phase}. With detailed balance,  energy fluctuations exist, but there is no mean transport of energy from any set of dynamical modes to any other. This is the case, for example, for truncated Euler equations.
We can therefore clearly understand why, as emphasized by Hopf and Kraichnan, the problem of viscous fluid flow differs radically from the inviscid limit. The question we address here is thus whether the energy flux across large scales is sufficiently weak for these scales to be treated as being in equilibrium, even though the overall system is far from equilibrium.\\

It has long been noted that a $k^2$ spectrum can also be observed in the case of decaying homogeneous isotropic turbulence. This system is an out of equilibrium transient and it is unlikely to be able to relate this $k^2$ spectrum to an equipartition mechanism as understood in equilibrium statistical mechanics. This results from a kinematic constraint. Provided that the velocity autocorrelation decays rapidly enough, the expansion of the energy spectrum for $k \rightarrow 0$ gives
\begin{equation}
E(k) = \frac{L}{4\pi^2} k^2 + \frac{I}{24\pi^2} k^4 + \cdots,
 \label{low k spectrum}
\end{equation}
where 
\begin{equation}
L=\int \langle {\bf u}({\bf x})\, {\bf u}({\bf x+r}) \rangle d^3r
\end{equation}
and 
\begin{equation}
I=-\int r^2 \langle {\bf u}({\bf x})\, {\bf u}({\bf x+r}) \rangle d^3r,
\end{equation}
are the Saffman and Loitsyansky integrals respectively \cite{saffman1967} (see \cite{davidson2015} for a review). Provided $L \neq 0$ (Saffman turbulence), a $k^2$ spectrum is obtained for decaying turbulence. Of course, decaying turbulence and forced turbulence in a statistically stationary regime are two different problems, and we can therefore think that the mechanisms generating a large-scale equipartition spectrum for turbulence forced at intermediate scales as observed in \cite{forster1977} have nothing to do with relation (\ref{low k spectrum}) which is purely kinematic. However, provided that the velocity autocorrelation decays rapidly enough, (\ref{low k spectrum}) should also apply to forced turbulence. This problem has been considered in great detail in \cite{hosking2023} where it has been shown that for a large class of forcings, $L$ is also invariant in forced turbulence. Starting from initial conditions with $L=0$ then could lead to a $k^4$ spectrum. A mechanism relying on turbulent diffusion of linear momentum has been proposed to explain the generation of an equipartition spectrum in forced turbulence.\\

This section is organized as follows: in subsection 2a, we present the first numerical evidence of the generation of an equipartition spectrum at large scales for forced turbulence at intermediate scale. Experimental results that support the conjecture of statistical equilibrium at scales larger than the forcing scale are then rewieved in subsection 2b. However, in subsection 2c, we report that deviations from the equipartition spectrum can occur depending on the type of forcing. In subsection 2d, we discuss these results and we establish the properties of the forcing that lead to the observation of energy equipartition among the large-scale modes. Truncated Euler equations are presented in subsection 2e and the correlation time of the large scales of a viscous turbulent flow are compared to the predictions of truncated Euler equations in subsection 2f.

\subsection{A first observation of equipartition at large scales}

It has been shown that when a three-dimensional flow is forced at a single wavelength $2\pi /k_f$ small compared to the size of the flow domain,  the statistics of the scales between the forcing scale and the domain size can be reasonably approximated as if they were in statistical equilibrium despite the fact that they are not isolated from the turbulent scales within the inertial range \cite{dallas2015}. The main results of this study are summarized below.

Flows should involve a large enough scale separation between the forcing scale $2\pi /k_f$ and the size of the periodic domain $2\pi$. In order to be able to resolve the turbulent flow in the inertial and dissipative ranges above $k_f$, the hyperviscous Navier-Stokes equation 
\begin{equation}
\frac{\partial {\bf u}}{\partial t}  + ({\bf u} \cdot \nabla) {\bf u} = - \nabla p + (-1)^{n+1} \nu_n \nabla^{2n} {\bf u} + {\bf f}
 \label{eq:hyperNS}
\end{equation}
is used, where ${\bf u}({\bf x}, t)$ is the solenoidal velocity field, $\nu_n$ is the constant hyperviscosity, $\bf f$ is the forcing function, which is described below and $p$ is the hydrodynamic pressure. Note that $n = 4$ is chosen here.

In the ideal case $\nu_n = 0$ and ${\bf f}=0$, Eq. \eqref{eq:hyperNS} conserves the kinetic energy $E = \frac{1}{2}\langle{\bf u}^2\rangle$ and the helicity $H = \langle{\bf u} \cdot {\bf \omega}\rangle$ where ${\bf \omega} = \nabla \times {\bf u}$ is the vorticity and angular brackets stand for spatial average. The level of helicity in the flow is given by the normalized helicity $-1 \leq \rho_H \equiv H / (\langle {\bf u}^2 \rangle \, \langle {\bf \omega}^2 \rangle )^{1/2} \leq 1$.
Eq. \eqref{eq:hyperNS} with $\nabla \cdot {\bf u} = 0$ is integrated using a standard pseudo-spectral code (see \cite{mininni2011hybrid} for the characteristics of the code).

Two types of forcing of the velocity field at  intermediate wavenumbers $k_f$ are used: a helical random forcing
 \begin{align}
 {\bf f}_H = f_0
  \{&[\cos(k_f y + \phi_y) + \sin(k_f z + \phi_z)]\hat x, \nonumber \\
    &[\cos(k_f z + \phi_z) + \sin(k_f x + \phi_x)]\hat y, \nonumber \\
    &[\cos(k_f x + \phi_x) + \sin(k_f y + \phi_y)]\hat z\} 
\label{eq:f_H}
 \end{align}
where ${\bf f}_H \cdot \nabla \times {\bf f}_H = k_f {\bf f}_H^2 > 0$ at each point in space and a non-helical random forcing
 \begin{align}
 {\bf f}_{NH} = f_0
  \{&[\sin(k_f y + \phi_y) + \sin(k_f z + \phi_z)]\hat x, \nonumber \\
    &[\sin(k_f z + \phi_z) + \sin(k_f x + \phi_x)]\hat y, \nonumber \\
    &[\sin(k_f x + \phi_x) + \sin(k_f y + \phi_y)]\hat z\}
 \end{align}
where $\langle {\bf f}_{NH} \cdot \nabla \times {\bf f}_{NH} \rangle = 0$. The phases $\phi_x$, $\phi_y$, $\phi_z$ are changed randomly at given correlation time scales $\tau_c$ small compared to the time scales of the large scales. Different scale separations $k_f = 10, 20, 40$ are considered, keeping the forcing velocity scale $u_f\propto (f_0/k_f)^{1/2}$ constant.  The value of $\nu_n$ is taken such that the Reynolds number, ${\rm Re} \equiv u_f k_f^{1-2n}/\nu_n$, is maintained in the range $6.\, 10^3 < {\rm Re} < 10^4$.

\begin{figure}[!ht]
   \includegraphics[width=0.5\textwidth]{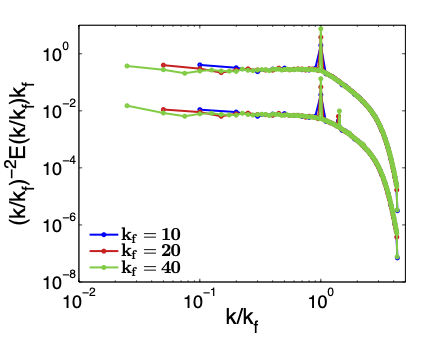}
   \includegraphics[width=0.5\textwidth]{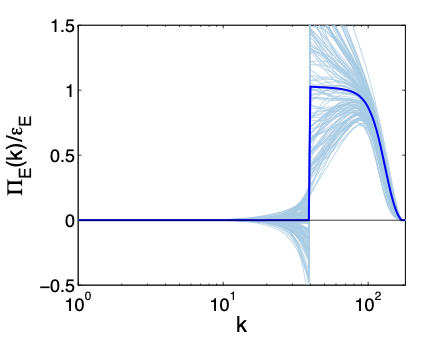}
 \caption{(left) Compensated $k^{-2}E(k)$ energy spectra for helical (top) and non-helical (bottom) flows. (right) $\Pi_E(k)/\epsilon_E$  for the helical flow with $k_f=40$. Thick blue lines represent the time-averaged values while thin gray lines the instantaneous values for various instants in time. Figure adapted from \cite{dallas2015}, "Copyright (2015) by The American Physical Society".}
 \label{spectreflux}
 \end{figure}

Figure \ref{spectreflux}(left) presents the energy spectra compensated with $k^{-2}$. Note that the energy spectra for different values of $k_f$ collapse since they are rescaled with $k/k_f$ and that $u_f$ is kept constant. The energy spectra for the helical and non-helical flows are shown with the non-helical spectra being shifted down for clarity. 

Figure \ref{spectreflux}(left)  displays a $E(k) \propto k^2$ scaling at low wavenumbers $k < k_f$ both for the helical and the non-helical flows. Similarly the collapsed helicity spectra (not shown) display the scaling $H(k) \propto k^4$. 
For isotropic turbulence, we have for the kinetic energy per unit mass, $V^2 = \int F({\vec k}) d^3k = \int E(k) dk$. Equipartition of energy, i.e. $F({\vec k})$ constant, therefore implies $E(k) \propto k^2$. 
Dimensional analysis then gives, 
\begin{equation}
E(k) \propto \frac{V^2 k^2}{{k_f}^3},
\end{equation}
as presently observed for wavenumbers smaller than $k_f$.

This scaling is in agreement with the absolute equilibria of the truncated Euler equations non-helical flows. In the case of helical flows, it has been shown by Kraichnan that a departure from the $k^2$ law is observed when $k$ is increased \cite{kraichnan1973}. This behavior is not observed in Figure \ref{spectreflux}(left) because, despite the fully helical forcing used (Eq. \ref{eq:f_H}), an insufficient amount of helicity is transferred to the large scales for the flow to become strongly helical. As a result, no singular deviation from the $H(k)\propto k^{4}$ and $E(k)\propto k^{2}$ power laws is observed due to the presence of helicity \cite{dallas2015}.

The energy flux $\Pi_E(k)$, normalized by the energy dissipation rate
$\epsilon_E = 2\nu \int_0^{\infty} k^{2n} E(k,t) dk$, is shown in Fig. \ref{spectreflux}(right) for the helical flow with $k_f = 40$.
For wavenumbers $k > k_f$, the time-averaged flux is positive and remains constant over the range $k_f < k < 2.5k_f$, indicating a forward energy cascade. In the range $k < k_f$, the time-averaged flux is zero, demonstrating that each large-scale mode receives, on average, the same amount of energy as it transfers to other modes. However, although the mean value of $\Pi_E(k)$ is zero, the instantaneous flux exhibits strong fluctuations of both signs. These arise because the power injected at $k = k_f$ varies in time, as both the velocity field and the forcing phases fluctuate. Consequently, local interactions are expected to generate significant flux fluctuations in the shells close to $k_f$. Moreover, Fig. \ref{spectreflux}(right) shows that while such fluctuations remain large within the direct cascade, they are progressively damped at smaller wavenumbers. Similar behavior is observed for the helicity flux (not shown).
The same behavior is observed for non helical forcing (not shown).

It should be noted that deviations from equipartition of energy do exist for the lowest wavenumber modes of the system $k \leq 2k_{min}$. These scales are slightly more energetic than the ones in the range $2k_{min} \leq k \leq k_f $. The amplitude of the deviation only weakly depends on scale separation.
Possible reasons for this behavior were mentioned in \cite{dallas2015}. First, at the largest scales close to the box size, the assumption of isotropy is not valid such that deviations from the isotropic result are expected. Second, a large scale instability could occur \cite{frisch1987,yakhot1987}. Such an instability can transfer energy directly from the forced and turbulent scales to the largest scales of the flow. Departure from equipartition stronger than the one reported here have been observed later for particular types of forcing of \cite{alexakis2019,ding2024}. This will be discussed in subsection 2c.

\subsection{Experimental observation of equipartition of energy among the large scale modes}

Following \cite{dallas2015}, several experiments were performed to check whether the scales larger than the forcing scale display an equipartition spectrum. They concern both three-dimensional hydrodynamic turbulence and wave turbulence. Some examples of wave turbulence share with three-dimensional hydrodynamic turbulence the characteristic that they involve a direct cascade of energy and no inverse cascade. Therefore, using similar arguments, we could expect that the scales larger than the forcing scale display energy equipartition. 

This was first shown for capillary wave turbulence \cite{michel2017}. Before describing the experiment, we sum up theoretical results. Waves on the surface $L^2$ of a horizontal fluid layer result from the restoring forces of gravity and surface tension. When viscous dissipation can be neglected, their dispersion relation in the deep layer limit is 
\begin{equation} \label{disp}
\omega^2 = g k + \left( \frac{\sigma}{\rho} \right) k^3,
\end{equation}
where $\omega$ is the angular frequency, $k$ is the wave number, $g$ is the acceleration of gravity, $\rho$ is the density of the fluid and $\sigma$ is the surface tension. The gravity and capillary terms are equal at a frequency $f_\mathrm{g.c.} = (2\pi)^{-1}(4\rho g^3/\sigma )^{1/4}$. The gravity dependent term can be neglected for frequencies large compared to $f_\mathrm{g.c.}$. We thereafter consider this regime of capillary waves.

If the waves are in thermal equilibrium at an effective temperature $T$, the isotropic energy spectral density per unit density and per unit surface $e_k^{(1D)}$ is given by $k_\mathrm{B} T / \rho L^2$ times the density of modes, 
\begin{equation} \label{energy_th}
e_k^{(1D)}= \frac{k_\mathrm{B}T k}{2\pi \rho}.
\end{equation}
$e_k^{(1D)}$ can be related to the power spectrum density of the surface elevation \textit{via} $S_\eta (k) =(\sigma / \rho)^{-1} k^{-2 }e_k^{(1D)} $,
\begin{equation}
S_\eta(k) = \left( \frac{\sigma}{\rho} \right)^{-1} \frac{ k_\mathrm{B}T}{2\pi \rho} k^{-1}.
\end{equation}
Using \eqref{disp}, the power spectrum density in the frequency domain $S_\eta(f) = 2\pi (\mathrm{d}k/\mathrm{d}\omega) S_\eta(k)$ is
\begin{equation}
S_\eta(f)=\frac{k_\mathrm{B}T}{3\sigma \pi f}. \label{eqth}
\end{equation}

In the absence of external forcing, $T$ is the room temperature. Thermal equilibrium of capillary waves has been studied by light scattering to check the validity of the dispersion relation at high frequencies \cite{Katyl1968} or to measure surface tension in the vicinity of the liquid-vapor critical point \cite{Bouchiat1969}.

When the waves are externally driven at a single frequency $f_\mathrm{inj}$, the system is no longer in thermal equilibrium but reaches a non-equilibrium steady-state. It displays a direct energy cascade characterized by \cite{Zakharov1967}

\begin{equation}
S_\eta(f) \sim  \epsilon^{1/2}\left(\frac{\sigma}{\rho}\right)^{1/6}f^{-17/6},
\label{weak_turbulence}
\end{equation}
where $\epsilon$ is the mean energy flux (per unit density and surface) cascading from large to small scales. The theory of weak turbulence predicts that even if such a system is very far from equilibrium, equipartition of energy \eqref{eqth}  should be observed in the range $f_\mathrm{g.c.}<f<f_\mathrm{inj}$ while the Kolmogorov type spectrum \eqref{weak_turbulence} is observed from $f_\mathrm{inj}$ up to a dissipative scale. Matching \eqref{eqth} and \eqref{weak_turbulence} at $f_\mathrm{inj}$ gives
\begin{equation} \label{temperature_vs_p}
k_\mathrm{B}T \sim \epsilon^{1/2}\sigma^{7/6}\rho^{-1/6}f_\mathrm{inj}^{-11/6},
\end{equation}
that was obtained within the weak turbulence framework in \cite{Balkovsky1995}. 

The experiment is conducted in a rectangular plastic vessel ($225 \times 180 \times 45~\mathrm{mm}$) filled with mercury to a depth of 30 mm. The fluid properties are $\rho = 13.5 \times 10^{3}~\mathrm{kg m^{-3}}$, $\sigma = 0.485~\mathrm{N m^{-1}}$, and $\nu = 1.15 \times 10^{-7}~\mathrm{m^{2} s^{-1}}$, yielding a gravity–capillary crossover frequency $f_{\mathrm{g.c.}} \simeq 16~\mathrm{Hz}$ and a wavelength of about $12~\mathrm{mm}$; bottom friction is therefore negligible.

Three wave makers (paddles of size $120 \times 80 \times 4~\mathrm{mm}$, immersed 10 mm below the free surface) are placed around a capacitive wave-height probe. The cavity size is chosen such that the lowest eigenfrequencies are a few hertz, strongly reducing the gravity-wave range. The probe wire diameter (0.35 mm) sets a high-frequency cutoff of a few hundred hertz.

Each wave maker is driven by a Brüel \& Kjær 4810 shaker, and its displacement $\xi_i(t)$ ($i=1,2,3$) is measured with a Brüel \& Kjær 4393 accelerometer. The $\xi_i(t)$ are independent realizations of band-limited random noise generated by an Agilent 33500B and filtered between $f_{\mathrm{inj}}$ and 200 Hz (SR 650). Several forcing amplitudes are used, with $f_{\mathrm{inj}}$ varied from 50 to 100 Hz. The surface elevation $\eta$ and accelerations $\ddot{\xi}_i$ are recorded using a National Instruments acquisition card; 50 Hz electrical noise is removed prior to analysis.

\begin{figure}[!ht]
   \includegraphics[width=0.5\textwidth]{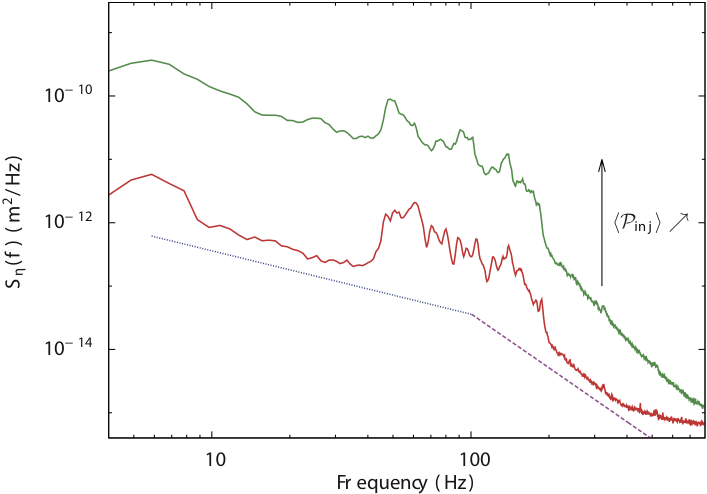}
 \caption{Power spectra of the surface wave height for two different forcing amplitudes with the same bandwidth (50 to 200 Hz). Power-laws theoretically predicted for the capillary thermal equilibrium spectrum (dotted line) and the direct energy cascade (dashed line). $\langle \mathcal{P}_\mathrm{inj}\rangle$ is the mean injected power. Figure adapted from \cite{michel2017}.}
 \label{spectres}
 \end{figure}

Two wave-height power spectra $S_\eta(f)$ are shown in Fig. \ref{spectres}. They are obtained at $f_{\mathrm{inj}} = 50~\mathrm{Hz}$ for two different forcing amplitudes. Between the gravity–capillary crossover frequency $f_{\mathrm{g.c.}}$ and the forcing frequency $f_{\mathrm{inj}}$, the spectra exhibit a $f^{-1}$ power law, characteristic of a thermal equilibrium state.

In some cases, the $f^{-1}$ scaling extends below $f_{\mathrm{g.c.}}$. As in gravity–capillary transitions observed in surface-wave turbulence driven at low frequencies, $f_{\mathrm{g.c.}}$ should be regarded as an approximate crossover frequency rather than a sharp boundary (see, \textit{e.g.}, Ref.~\cite{Falcon2007}). At higher frequencies, from about 200 Hz up to the dissipative scale, a second self-similar regime is observed. Because the energy flux is not conserved along this cascade, owing to dissipation and to the high-frequency cutoff of the measurement system, the measured spectra are steeper than the weak-turbulence prediction \eqref{weak_turbulence}. This high-frequency cascade has been extensively studied over the past two decades and is not discussed further here.

For each spectrum, an effective temperature is computed by integrating the wave-height power spectrum over the thermal range:
\begin{equation}
T_\mathrm{eff}= \frac{3\pi \sigma }{k_\mathrm{B}} \frac{\int_{f_\mathrm{g.c.}}^{0.85 f_\mathrm{inj}} S_\eta(f) \mathrm{d}f}{\ln (0.85 f_\mathrm{inj}/f_\mathrm{g.c.})}
\label{Teff}
\end{equation}
The power spectrum $S_\eta (f)$ is fitted by a power-law $\mathrm{cst} \times f^\alpha$ in the same frequency range ($f_\mathrm{g.c.}$ to $0.85 f_\mathrm{inj}$) for 10 different forcing amplitudes on different frequency ranges.  The values of $\alpha$ stand close to -1, as expected for a thermal equilibrium range \cite{michel2017}. 
It has been also shown that the effective temperature is governed by the injected power, $T_\mathrm{eff} \propto \langle \mathcal{P}_\mathrm{inj}\rangle$ on three decades. 

In conclusion, it has been shown that scales larger than the forcing scale are in equipartition and that their effective temperature, i. e. the average energy per mode, is determined by the energy flux within the energy cascade toward small scales for scales smaller than the forcing scale, or equivalently the injected or dissipated power by the system.\\

Other studies of wave turbulence in different systems have reached a similar conclusion. One is related to bending waves in a thin elastic plate generated by a sinusoidal forcing at frequency $f_0$ \cite{miquel2021}. This system also involves a direct energy cascade up to a cut-off frequency above which most of the energy is dissipated although dissipation also affects the inertial range. Using a high enough forcing frequency $f_0$, a flat frequency spectrum for the local plate velocity has been observed for $f < f_0$. Using the dispersion relation for bending waves, this corresponds to a spatial energy spectrum proportional to $k$, i. e. to an equipartition spectrum. A difference from the previous system with capillary waves was the absence of an inertial range due to the value of $f_0$ within the dissipative range. The flat equipartition spectrum at low frequency therefore directly transitions to a dissipative spectrum with an exponential decay. Despite this difference, the effective temperature for long waves was also determined by the injected or dissipated power.\\

Another study of wave turbulence that displayed equipartition of energy for modes with wavelength larger than the forcing scale is related to hydroelastic wave turbulence, i. e. waves in a thin elastic layer overlying an incompressible fluid \cite{vernet2025}. In the range of parameters studied by the authors, the tension applied to the sheet provides the dominant restoring force, such that the dispersion relation is approximately the one of capillary waves. Therefore the equilibrium spatial (respectively temporal) spectrum is a $k^{-1}$ (respectively $f^{-1}$) power law as for capillary waves. Both spectra are experimentally observed. As in the previous study on bending waves in a thin elastic plate, the equilibrium large scale spectrum directly transitions to a dissipative spectrum at higher wavenumber or frequency.  

Three different experiments on wave turbulence have therefore shown that the modes of the system with wavelength larger than the forcing scale can display an equipartition spectrum at equilibrium although the scales smaller than the forcing scale experience an energy flux toward smaller scales and dissipation of energy. The effective temperature of the large scales is determined by the amount of dissipated power. The three systems involve no inverse cascade toward large scales and a periodic forcing or a noisy periodic forcing at a much smaller scale than the ones of the thermalized modes. \\

Equipartition of energy among the large scales has been also observed experimentally in three-dimensional hydrodynamic turbulence \cite{gorce2022}.
The experiment consists of a turbulent flow of water generated in a parallelepipedic container $32 \times 32 \times 22$ ${\rm cm}^3$ by the random motion of centimetric magnetic particles forced by a time dependent magnetic field. This provides a clever way to generate a fairly homogeneous and isotropic turbulent flow with a small mean flow component. The forcing scale is estimated to be $5$ cm which gives a limited scale separation with the container size. The velocity field is measured using particle image velocimetry in the horizontal plane and one-dimensional spectra are computed. They display an inertial range in agreement with Kolmogorov law on less than a decade and become flat at low wavenumber. The three-dimensional spectrum is computed from them assuming isotropy. It displays an equilibrium spectrum proportional to $k^2$ on less than a decade at the largest scales. As for experiments on wave turbulence, it is shown that the energy flux determines the effective temperature, i. e. the energy per mode in the equilibrium large scale range. It is also shown that the energy flux vanishes on average in the range of scales larger than the forcing scale although strong fluctuations are observed in agreement with the numerical simulations displayed in subsection 2a. It should be noted that the forcing in this last experiment differs qualitatively from the periodic or noisy periodic forcings of the previous experiments. Here, the flow is forced by kicks provided by the random motion of particles. This forcing is localized in real space as the one used in some numerical simulations \cite{ding2024} but probably not very narrow in Fourrier space. It would have been interesting to estimate the spectral content of the forcing by measuring the flow in the linear regime at low forcing amplitude. In any case, an equipartition spectrum was observed.

\subsection{Departures from Equipartition}\label{sec:departure}

As noted in subsection 2a, it has been observed that the largest scales may exhibit an excess of energy relative to equipartition \cite{dallas2015}. Much more important deviations from equipartition at large scales were investigated in detail in Ref.~\cite{alexakis2019}, where it was shown that the nature of the forcing plays a crucial role in determining the behavior of large-scale modes. In particular, a spectrally dense forcing, defined as one that excites all modes within a finite-width spherical shell, and possessing long temporal correlations, can lead to strong departures from the expected $k^2$ energy spectrum. By contrast, a spectrally sparse forcing, acting on only a few modes and characterized by short correlation times, was shown to recover the thermal equilibrium spectrum.

These deviations from equipartition were analyzed in terms of several diagnostics, including:
\begin{itemize}
\item the number of contributing triadic interactions;
\item the spectrum of the nonlinear term;
\item the amplitude of interactions and the associated interscale energy fluxes;
\item the transfer function between different wavenumber shells.
\end{itemize}

It was found that spectrally dense forcing promotes a large number of triadic interactions that couple large-scale modes with pairs of forced modes, leading to an anomalously large energy injection at large scales. This excess energy is subsequently transferred back toward smaller scales through both self-interactions among large-scale modes and interactions with turbulent small-scale modes.

The resulting picture is one of nonlocal interactions between two distinct reservoirs: the large-scale modes, and a second reservoir composed of small-scale, forced, turbulent, and dissipative modes that remains far from equilibrium.

The authors concluded that when the energy transfer from the forced modes to the large scales is weak, as in the case of spectrally sparse forcing, the large-scale energy spectrum remains close to thermal equilibrium. In this regime, non-local interactions in Fourier space merely set an effective global energy (or temperature) for the system. Conversely, when long-range interactions dominate, as with spectrally dense forcing, large-scale self-interactions are too slow to restore equilibrium. This leads to sustained deviations from equipartition and the emergence of modified large-scale spectral scalings.\\

Departures from equipartition were further investigated in Ref.~\cite{ding2024}. The authors emphasized that, for the specific forcings considered, namely forcings that are $\delta$-correlated in time with an exponentially decaying spatial autocorrelation, or forcings with a $\delta(k-k_f)$ power injection rate in spectral space, the Kármán–Howarth equation implies that the third-order structure function of velocity increments decays slowly, as $r^{-2}$, in the large-scale limit. This behavior indicates that a small but finite influence of the forcing persists at large scales. The authors indeed observed strong deviations from equipartition, with spectra scaling as $k^{1/2}$ and $k^{3/2}$ for the two forcings considered, to be compared with the more extreme $k^0$ scaling reported in Ref.~\cite{alexakis2019} for spectrally dense forcing. However, when a six-mode spectrally sparse forcing was employed, an equipartition spectrum was recovered, despite the third-order structure function exhibiting the same slow decay as in the previous cases.\\

\subsection{Characteristics of the forcing for which equipartition of energy is expected at large scales}

When departures from equipartition are observed, it is natural to ask how relation~(\ref{low k spectrum}) breaks down. In particular, do coherent structures emerge at the largest scales? If so, can their influence be suppressed, for instance by mimicking the effect of physical boundaries as in laboratory experiments? Moreover, are the observed power laws with noninteger exponents genuine asymptotic scalings, or do they arise from finite-size effects?

In the above-cited studies, the authors often relate departure from equipartition with departure from statistical equilibrium. However, all investigations agree that the mean energy flux across large scales vanishes. This raises the question of what constitutes nonequilibrium behavior in a system with zero mean energy flux. We emphasize that a departure from equipartition does not necessarily imply a departure from equilibrium. The equipartition theorem relies on several assumptions, and well-known examples exist in equilibrium statistical mechanics of systems that are at equilibrium yet do not exhibit equipartition. 

Another important theoretical issue arises in the discussion of departures from equipartition. While it is perhaps unsurprising that deviations occur when the forcing is designed such that large-scale modes are directly coupled to many triadic interactions that involve two externally forced modes, it should be recalled that the first, and so far only analytical prediction of equipartition at large scales for Navier–Stokes flows forced at intermediate scales is the calculation of Ref.~\cite{forster1977} (Model C). Although the forcing employed in that paper is not spectrally sparse, it generates an equipartition spectrum at large scales. According to the analysis of reference \cite{alexakis2019}, deviations from equipartition arise from the existence of triads in which the beating of two forcing wave vectors directly excites a large-scale mode. This configuration is only possible if, for a fixed domain, the forcing is made increasingly spectrally dense, so that the angle between two neighboring forcing wave vectors, with magnitude on the order of $k_f$, becomes sufficiently small. Conversely, for a given forcing, if the domain size is increased enough, the direct excitation of the largest scales by two forcing wave vectors is no longer possible. More precisely, if we define $\Delta k$ as the minimum value of $\vert {\bf k}_{fi} - {\bf k}_{fj} \vert$ where the ${\bf k}_{fi}$'s are the forcing wave vectors of magnitude close to $k_f$, equipartition at large scale will be observed provided $\Delta k \,L \gg 1$ in addition to $k_f\, L \gg 1$. In that case, we can expect a universal behavior of the large scales in the form of an equipartition spectrum for $k \ll \Delta k$.

\subsection{Truncated Euler equations}

We have diagnosed so far the possible equilibrium of large scales by considering spatial spectra or equivalently spatial correlations evaluated at a fixed time. In a turbulent flow, one can also consider temporal correlations of the (Eulerian) velocity evaluated at a fixed location, or spatio-temporal correlations involving two times and two points. If we assume that the forcing weakly affects the large scales, such that their dynamics is primarily due to the nonlinear interaction between them, we may hope to model the dynamics of the large scales using the Euler equation truncated at the forcing scale or below. The truncated Euler equations can therefore provide a tool to compute the dynamics of large scales at equilibrium and to compare with the results of direct simulations or 
experiments. 

Let us first recall some results about the truncated Euler equations (TEE). 
The three-dimensional, spatially-periodic, forced Navier-Stokes equations, can be solved numerically using $2/3$ de-aliased Fourier spectral methods~\cite{gottlieb1977}. In this framework, the velocity field ${\bf v}(t,{\bf x})$ is represented by its discrete Fourier transform:
\[
{\bf v}(t,{\bf x}) = \sum \hat{{\bf v}}(t,{\bf k}) e^{i {\bf k} \cdot {\bf x}},
\]
and evolves according to the Navier-Stokes equations (assuming unit density):
\begin{eqnarray}
\partial_t {\bf v} + ({\bf v} \cdot \nabla) {\bf v} &=& - \nabla p + \nu \nabla^2 {\bf v} +  {\bf f}, \nonumber \\
\nabla \cdot {\bf v} &=& 0,
\label{eq_sns}
\end{eqnarray}
where ${\bf f}$ is force per unit mass.

Applying a Galerkin truncation to Eq.~\eqref{eq_sns}-i.e., setting $\hat{{\bf v}}(t,{\bf k}) = 0$ for $k_\alpha k_\alpha \ge k_{\rm max}^2$ (where $k_{\rm max}$ is one-third the resolution)-yields a finite system of ordinary differential equations for the Fourier modes $\hat{{\bf v}}(t,{\bf k})$, where ${\bf k}$ is a three-dimensional vector of integers $(k_1,k_2,k_3)$. In the inviscid and unforced case ($\nu = 0, f=0$), the system reduces to the TEE
\begin{equation}
\partial_t \hat{v}_\alpha(t,{\bf k}) = -\frac{i}{2} \mathcal{P}_{\alpha \beta \gamma}({\bf k}) \sum_{\bf p} \hat{v}_\beta(t,{\bf p}) \hat{v}_\gamma(t,{\bf k}-{\bf p}),
\label{tee}
\end{equation}
where $\mathcal{P}_{\alpha \beta \gamma}({\bf k}) = k_\beta P_{\alpha \gamma}({\bf k}) + k_\gamma P_{\alpha \beta}({\bf k})$, with $P_{\alpha \beta} = \delta_{\alpha \beta} - k_\alpha k_\beta / k^2$. The convolution sum is truncated to zero for $k_\alpha k_\alpha \ge k_{\rm max}^2$.

The TEE are time-reversible and conserve the total kinetic energy
\[
E = \sum_k E(t,k),
\]
where the energy spectrum $E(t,k)$ is defined by summing $|\hat{{\bf v}}(t,{\bf k}')|^2$ over spherical shells of width $\Delta k = 1$:
\begin{equation}
E(t,k) = \frac{1}{2} \sum_{k - \Delta k/2 < |{\bf k}'| < k + \Delta k/2} |\hat{{\bf v}}({\bf k}',t)|^2.
\label{eq_energy}
\end{equation}

The nonlinear term on the right-hand side of Eq.~\eqref{tee},
\[
\mathcal{N}_\alpha({\bf k};\hat{{\bf v}}) = -\frac{i}{2} \mathcal{P}_{\alpha \beta \gamma}({\bf k}) \sum_{\bf p} \hat{v}_\beta(t,{\bf p}) \hat{v}_\gamma(t,{\bf k}-{\bf p}),
\]
satisfies the Liouville property:
\[
\sum_{\hat{{\bf v}}({\bf k})} \frac{\partial \mathcal{N}_\alpha({\bf k};\hat{{\bf v}})}{\partial \hat{{\bf v}}({\bf k})} = 0.
\]

In the absolute equilibrium of the deterministic, truncated Euler system, a standard statistical mechanics argument (see e.g. Refs.~\cite{lee1952,kraichnan1973,orszag1977}) shows that the microcanonical distribution
\begin{equation}
P_{\rm mc}[\hat{{\bf v}}] = Z_{\rm mc}^{-1} \delta(E(\hat{{\bf v}}) - E),
\label{eq:Pmicro}
\end{equation}
can be approximated, for large numbers of degrees of freedom (scaling with the cube of the resolution), by the canonical distribution:
\begin{equation}
P_{\rm sta}[\hat{{\bf v}}] = Z_{\rm c}^{-1} e^{-\beta E},
\label{eq:pstationary}
\end{equation}
where $Z_{\rm mc}$ and $Z_{\rm c}$ are normalization constants.

This result can be seen directly by introducing the Liouville equation for the probability density $\mathbb{P}\left[\{ \hat{{\bf v}}_{k}, \hat{{\bf v}}_{k}^*\}_{0\leq k_1 \leq k_{\rm max}} \right]$:
\begin{equation}
\frac{\partial \mathbb{P}}{\partial t} = \sum_{\hat{{\bf v}}({\bf k}); 0 \leq k_1 \leq k_{\rm max}} \frac{\partial}{\partial \hat{{\bf v}}({\bf k})} \left[ -\mathcal{N}_k(\hat{u}) \mathbb{P} \right] + \text{c.c.},
\label{eq:Liou}
\end{equation}
where $\hat{{\bf v}}_{k}^* = \hat{{\bf v}}_{-k}$ is treated as an independent variable, and "c.c." denotes the complex conjugate.

Due to energy conservation, the stationary distribution \eqref{eq:pstationary} is a solution of Eq.~\eqref{eq:Liou}.\\

\subsection{Correlation time of large scales}

The truncated Euler equations (TEE) have been used to compute time-dependent cross-correlation functions between Fourier modes at different wavenumbers, from which a decorrelation time as a function of $k$ can be extracted. As shown in Ref.~\cite{cichowlas2005}, the decorrelation time $\tau(k)$ of TEE modes at wavenumber $k$ exhibits a short-time scaling $\tau(k) \sim (u_{\rm rms} k)^{-1}$, corresponding to a parabolic behavior of the correlation function at early times. At longer times and for lower wavenumbers, the correlation function crosses over to an exponential decay~\cite{cichowlas2005b,gosteva2025}, characterized by a timescale $\tau(k) \sim (\nu_{\rm eff} k^2)^{-1}$, where $\nu_{\rm eff} \sim u_{\rm rms}/k_{\rm max}$ and $u_{\rm rms} = \sqrt{2E}$ is the root-mean-square velocity.

The influence of helicity on the correlation time of large-scale modes in turbulent flows was investigated numerically in Ref.~\cite{cameron2017}. Direct numerical simulations (DNS) were performed for both Taylor–Green symmetric non-helical flows and fully periodic helical flows. The analysis encompassed the probability density functions, energy spectra, auto-correlation functions, and correlation times, allowing for a detailed comparison between the two classes of flows.

An analytical expression for the correlation time in the absolute equilibrium of helical flows was derived. In the strongly helical regime, this leads to a short-time decorrelation scaling of the form $\tau(k) \sim H^{-1/2} k^{-1/2}$, in contrast with the $\tau(k) \sim E^{-1/2} k^{-1}$ scaling obtained for weakly helical flows.

For Navier–Stokes simulations with Taylor–Green forcing (non-helical and characterized by a large separation between forcing and box scales), the large-scale modes were found to closely follow the predictions of absolute equilibrium. In particular, the correlation time displayed a $\tau(k) \sim E^{-1/2} k^{-1}$ scaling. While fully periodic helical flows also exhibited qualitative similarities with absolute equilibrium behavior, their largest scales showed more pronounced deviations from theoretical expectations.

Ref.~\cite{cameron2017} concluded that DNS of the truncated Euler equations exhibit excellent agreement with absolute equilibrium theory, with both predicted power laws for the correlation time in helical and non-helical cases being accurately recovered.

The situation is, however, less clear for the large scales of forced Navier–Stokes flows. Simulations employing the non-helical Taylor–Green geometry, which allows for greater scale separation, showed strong agreement with absolute equilibrium predictions: the large-scale velocity modes displayed nearly Gaussian statistics with variances consistent with equipartition, and the correlation time followed a power-law behavior compatible with a $k^{-1}$ scaling over nearly two decades in wavenumber.

By contrast, simulations of fully periodic flows exhibited weaker agreement. Although the large-scale energy spectra matched absolute equilibrium predictions over a limited range of scales, significant deviations were observed at the largest wavenumbers, which were found to be both more energetic and more helical than predicted.

Moreover, while the correlation time in helical flows was observed to decrease with increasing helicity, as anticipated from thermalization arguments, the measured correlation times deviated substantially from the $k^{-1}$ and $k^{-1/2}$ scaling laws. The physical origin of these deviations, and more generally the departure of the largest scales in forced Navier–Stokes flows from absolute equilibrium predictions, remains an open question.

%%%%%%%%%%%%%%%%%%%%%%%%%%%%%%%%%%%%%%%%%%%%%%%%%%%%%%%%%%%%%%%%%%%%%%%%%%%%%%%%%%%%%%%%%%%%%%%%%%%%%%%%%%%%%%%%%%%%%%%%
\section{Dynamical and statistical properties of large scales in two-dimensional turbulent Kolmogorov flows}
%%%%%%%%%%%%%%%%%%%%%%%%%%%%%%%%%%%%%%%%%%%%%%%%%%%%%%%%%%%%%%%%%%%%%%%%%%%%%%%%%%%%%%%%%%%%%%%%%%%%%%%%%%%%%%%%%%%%%%%%

Two-dimensional turbulence displays behavior fundamentally distinct from that of three-dimensional flows \cite{Tabeling2002,Boffetta2012}. When forcing is applied at a scale $l_f$, energy cascades inversely toward large scales, while enstrophy is transferred to small scales \cite{kraichnan1967}. This inverse energy cascade leads to the emergence of large-scale coherent structures (vortices, jets), as commonly observed in geophysical flows.

In realistic situations, the two-dimensional Navier–Stokes equation includes a linear friction term $- {\bf v}/\tau$, which induces dissipation at all scales. For an inverse energy flux per unit mass $\epsilon$, the energy spectrum peaks at a scale $l_I \propto (\tau^3 \epsilon)^{1/2}$, which effectively marks the arrest of the inverse cascade. If $l_I < L$, where $L$ is the domain size, a $k^{- 5/3}$ inverse cascade is observed \cite{sommeria1986,paret1997}. If $l_I > L$, energy accumulates at the largest accessible scale, typically forming vortical structures. This condensed state was predicted by Kraichnan using truncated Euler equations. As in three dimensions, the resulting dynamical system satisfies Liouville’s theorem and conserves energy; it also conserves enstrophy. Equilibrium statistical mechanics then predicts different regimes depending on initial conditions, including a condensed state in which most kinetic energy resides in the largest-scale mode.

Large-scale coherent structures can also be derived from equilibrium statistical mechanics applied to the two-dimensional Euler equation in real space. In Onsager’s point-vortex model \cite{onsager1949}, the dynamics form a Hamiltonian system amenable to statistical treatment. For continuous vorticity fields, the Miller–Robert–Sommeria theory \cite{miller1990,robert1991} accounts for the full set of invariants and yields entropy-maximizing states, enabling predictions of mean flows and transitions between flow geometries.

Fourier-space and real-space approaches share qualitative agreement but differ in scope \cite{kraichnan1975}. Both predict large-scale organization associated with negative temperature states. However, Fourier truncation retains only energy and enstrophy, whereas real-space methods incorporate all invariants. The two approaches are therefore complementary: real space provides insight into mean-flow geometry, while Fourier space describes fluctuation spectra. In both cases, neglecting forcing and dissipation is justified a posteriori, taking into account the qualitative agreement with observations. Notably, unlike in three-dimensional turbulence, the inverse cascade implies a nonzero mean energy flux across large scales.

Finally, studies of the condensed regime using quasi-linear theory  \cite{laurie2014,frishman2018,kolokolov2020,doludenko2021}
report a large-scale spectrum, with a $k^{- 3}$ or a $k^{-5}$ scaling that have 
received some confirmation from experiments and numerical simulations \cite{chertkov2007,Xia2009,laurie2014,zhu2024}.
These studies however use explicitly a balance of the injected energy rate with the dissipation and are thus out-of-equilibrium predictions. In a more recent study \cite{vankan2025} a transition from equilibrium spectra to quasi-linear predictions was observed as $Re$ was varied. To which extent this could validate the statistical approach is an open question.

This section is mostly devoted to the study of the transitions between the turbulent regimes observed as the large scale damping of the flow is reduced such that there are transitions from a turbulent regime not affected by the boundaries of the flow domain to a condensed state. We have identified two successive transitions. The first one is characterized by the probability density of the velocity that transitions from Gaussian to bimodal. The bimodality is related to random reversals of the large scale flow. It becomes more pronounced and the waiting time between successive reversals becomes longer when the damping rate is reduced. At a second transition, reversals stop and the flow is in the condensed regime. We have observed this behavior both experimentally (subsection 3a) and in direct numerical simulations of the two-dimensional Navier-Stokes equation with viscous drag (subsection 3b). We show in subsection 3c that truncated Euler equations reproduce qualitatively the transitions between these different regimes even though they are out of equilibrium.

%%%%%%%%%%%%%%%%%
\subsection{Experimental results}
%%%%%%%%%%%%%%%%%

\begin{figure}
  \centerline{\includegraphics[width=\textwidth]{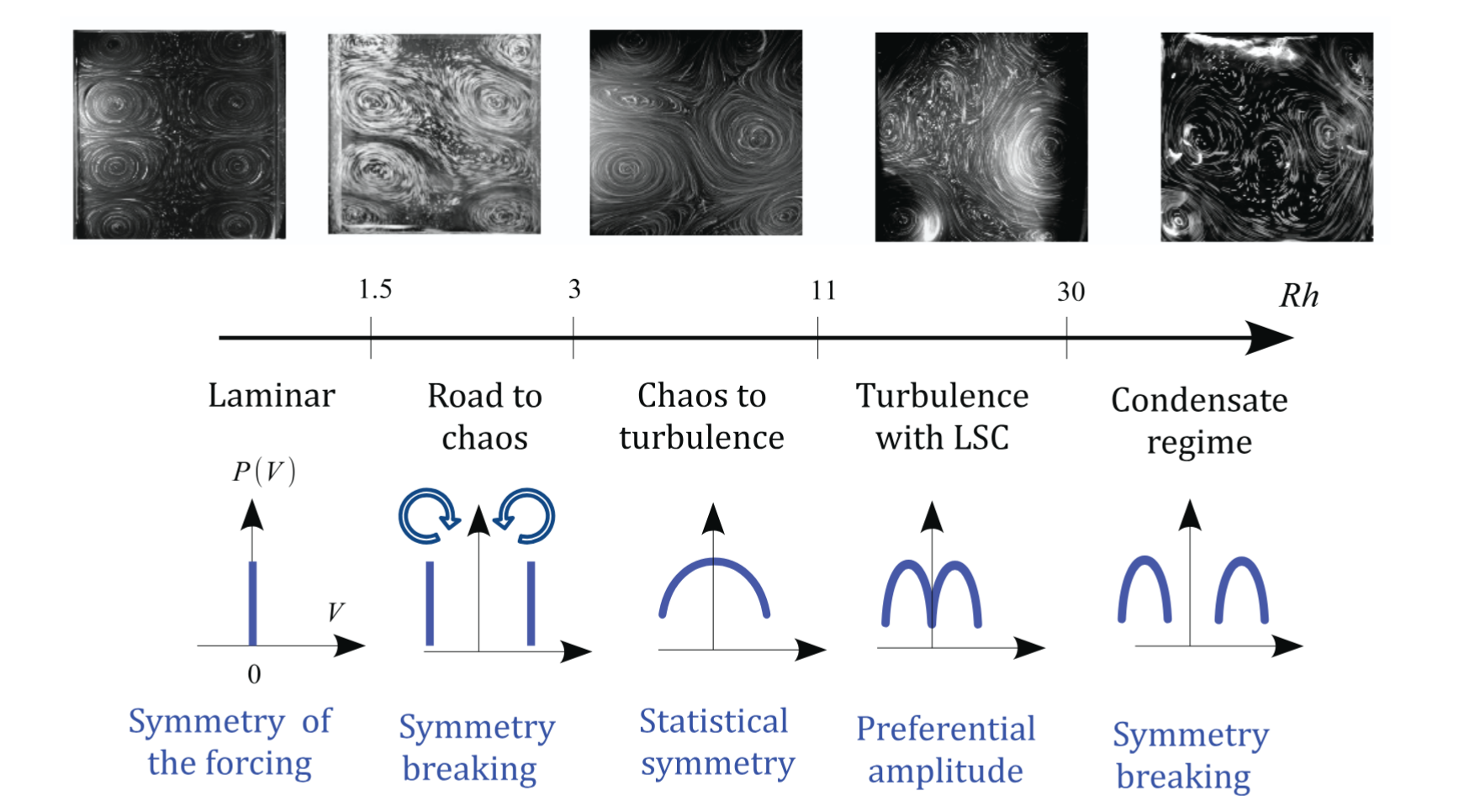}
   }
   % /mnt/6a509136-da71-4a58-b0d8-02e60d6215a1/04_Programmes/Prog_Matlab/openpiv-matlab-master/openpiv-matlab-master_Poiseuille'
  \caption{Top: pictures of the flow as a function of ${\rm Rh}$. From left to right: laminar flow, first bifurcation, chaotic
flow, turbulent flow with moderate large-scale flow, turbulent flow with strong large-scale flow (condensate). Bottom: sketch of
the PDF of large-scale circulation. Symmetry breaking indicates that depending on the initial conditions one of the two states
(i.e. one of the two peaks of the PDF) will be observed. Figure from \cite{michel2016}.}
\label{fig:Rh}
\end{figure}

Quasi 2-D flows forced periodically in space, also known as Kolmogorov flows, have been studied experimentally to investigate the formation of large scales \cite{Bondarenko1979, tabeling1987} and 2D turbulence  \cite{paret1997,Boffetta2012,Xia2009}. In \cite{michel2016} we investigated the sequence of bifurcations that lead to the formation of a large scale flow in a nearly 2D flow forced periodically at smaller scale. The experimental set-up consists in a layer of liquid metal, Gallinstan, of thickness $h=2$ cm and horizontal length $L=12$ cm. It is put into a vertical magnetic field up to $B=0.1$ T. An array of $2 \times 4$ electrodes at the bottom of the layer inject a current up to $I = 200$ A. The resulting Lorentz force, when small enough, drives a  stationary laminar flow made of $8$ vortices as displayed in fig. \ref{fig:Rh} top left.

In this experiment, the control parameter is ${\rm Rh}=\sqrt{\frac{I h}{B \sigma \nu L^2}}$ that compares the inertial term to the friction term due to the magnetic field. $\nu$ and $\sigma$ are the kinematic viscosity and the electrical conductivity of the liquid metal. The Reynolds number ${\rm Re}$ is much larger, ${\rm Re}/{\rm Rh} \simeq 10^4$. As depicted in fig \ref{fig:Rh}, a series of bifurcations lead the system to a chaotic regime for ${\rm Rh}\le 3$.

\begin{figure}
  \centerline{\includegraphics[width=0.5\textwidth]{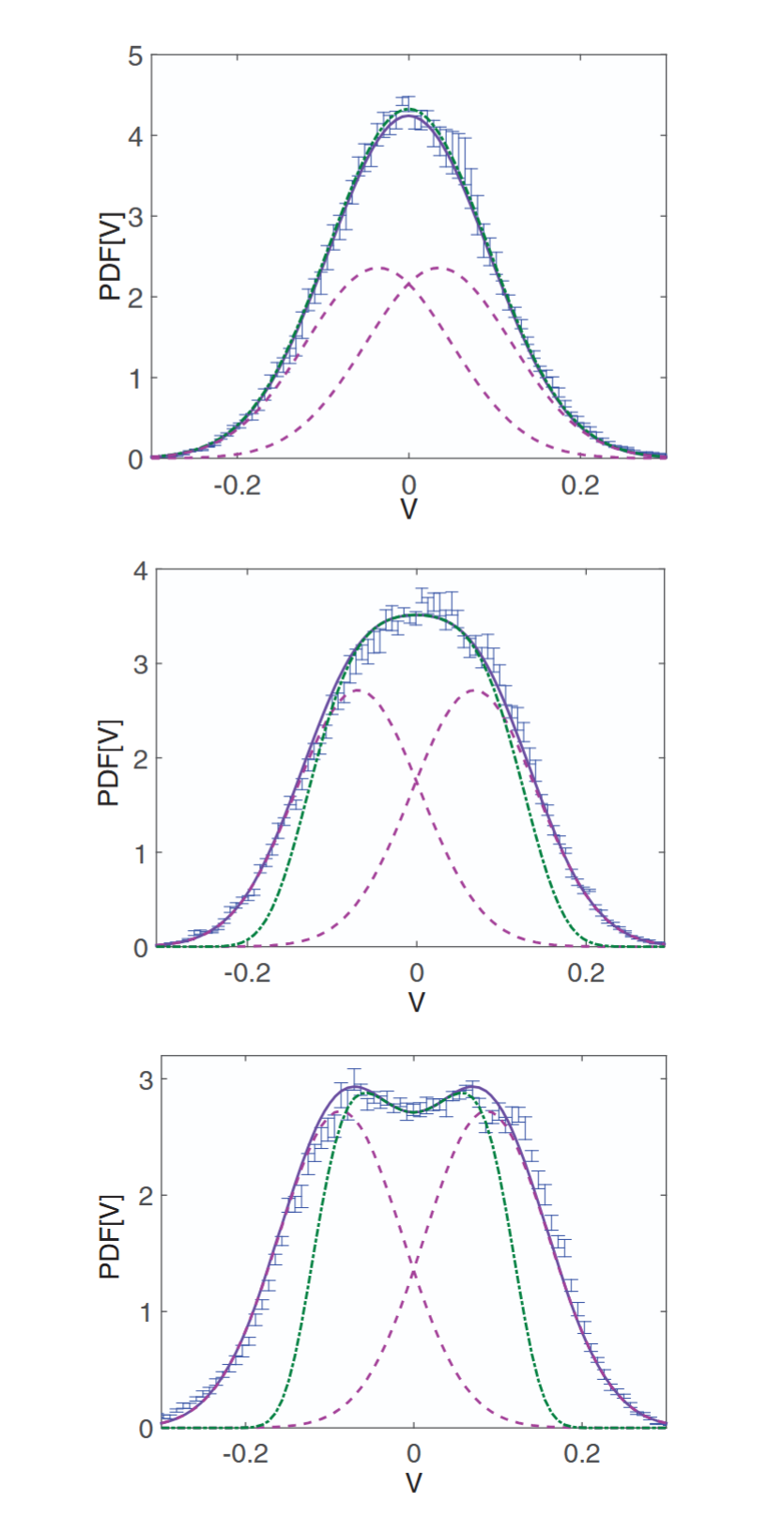}
   }
   % /mnt/6a509136-da71-4a58-b0d8-02e60d6215a1/04_Programmes/Prog_Matlab/openpiv-matlab-master/openpiv-matlab-master_Poiseuille'
  \caption{PDF of the amplitude of the large scale
velocity for ${\rm Rh} = 12, 14, 20$ (from top to bottom). Symbols
are experimental datas. The blue continuous curve is the fit with 
the sum of the two Gaussians displayed with magenta dashed
curves. The green curve is  fit using a Landau type free energy that does not capture correctly the tails of the distribution. Figure from \cite{michel2016}.}
\label{fig:pdf}
\end{figure}

For ${\rm Rh}$ slightly larger than $3$, the large scale circulation is observed to fluctuate around zero and its distribution is close to a Gaussian, with its maximum at zero. Further increase of ${\rm Rh}$ leads to a distribution that departs from a Gaussian and evolves towards a bimodal distribution, as displayed in fig. \ref{fig:pdf}.

A model of this evolution is provided by the sum of two Gaussian of width $\Delta$ and centered at $\pm dX$. A fit of the data with this function is displayed in blue and is in good agreement with the measured distribution.
The model parameters are as follows: $\Delta$ remains constant and $dX$ is zero for ${\rm Rh} \le {\rm Rh}_c \simeq 11$ and behaves as $\sqrt{{\rm Rh}-{\rm Rh}_c}$ for larger ${\rm Rh}$.

In addition, we note that the dynamics of the large scale consists in sign changes and can be described as random reversals between two states. The time series display a low frequency spectrum of the form $S(f)\propto f^{-\alpha}, $ with $0<\alpha<2$, which is usually named $1/f$ noise.  We have shown that this spectrum is associated to the long durations during which the system remains in the same state.  More precisely, we have shown in \cite{Herault2015a, Herault2015b}, that the duration during which a state is maintained is distributed as a power-law and that this power-law determines the exponent $\alpha$ of the spectrum. The origin of the power-law distribution of the duration between sign changes remains unknown and it would be interesting to understand it.

When ${\rm Rh}$ is increased above ${\rm Rh} \simeq 11$, we therefore observe a continuous transition from a turbulent regime with a Gaussian PDF of the velocity field, to a bimodal regime with two most probable opposite values of the large scale circulation. When ${\rm Rh}$ is increased in the range $20 < {\rm Rh} < 50$, the bimodality of the PDF becomes more and more pronounced and reversals of the large scale circulation become apparent on the direct recordings (see also Fig. \ref{direct recordings}). This reversal phenomenon of a large scale field is well-known, for instance in the case of a magnetic field generated by a dynamo process or for the large scale velocity in turbulent convection \cite{gallet2012}.

When ${\rm Rh}$ is increased further, the mean waiting time between successive reversals increases and the system finally stays in one polarity of the large scale circulation. We reach the so-called condensed state of two-dimensional turbulence where most of the kinetic energy of the flow is in the largest scale modes. 

This experiment therefore provides a nice example of transitions or bifurcations between qualitatively different turbulent flow regimes. A first transition from a turbulent flow with a Gaussian PDF to a bimodal PDF related to the dynamics of the large scale circulation, and a second transition to a condensed state. We will show in subsection 2c how this type of transitions can be modeled.  

%%%%%%%%%%%%%%%%%%
\subsection{Direct numerical simulations of the two-dimensional Navier-Stokes equation with drag}
%%%%%%%%%%%%%%%%%%

The dimensionless two-dimensional Navier-Stokes equations (2D NSE) for an incompressible flow, with velocity field ${\bf u} = \nabla \times \psi$, read:
\begin{equation}
\frac{\partial \psi}{\partial t} - {\nabla^{-2}}\{\psi,\nabla^2\psi\} = -\frac{1}{{\rm Rh}}\,\psi + \frac{1}{{\rm Re}}\, \nabla^2 \psi + f_\psi \, ,
\label{eq:NS_incom}
\end{equation}
where $\psi(x,y,t)$ is the stream function, and $\{f,g\} = \partial_x f \partial_y g - \partial_x g \partial_y f$ denotes the usual Poisson bracket. The first term on the right-hand side models bottom friction, while the forcing term is spatially periodic. It is given explicitly here by $f_\psi = \frac{1}{144} \sin(6x)\sin(6y)$.

To account for confinement effects, free-slip boundary conditions are employed. This allows for a Fourier expansion of the stream function of the form:
\begin{equation}
\psi(x,y) = \sum_{m,n} \hat{\psi}_{m,n} \sin(m x)\sin(n y).
\label{eq:sfbasis}
\end{equation}

The two non-dimensional parameters governing the system are the Reynolds number, ${\rm Re} = UL/\nu$, and the friction parameter ${\rm Rh} = \tau U / L$, where $U$ is a characteristic velocity, $L$ the domain size, $\nu$ the kinematic viscosity, and $1/\tau$ the damping rate associated with bottom friction. The system is non-dimensionalized using $L$ and $U$.

\begin{figure}
\vspace {-4 cm}
\centering
\includegraphics[width=0.48\textwidth]{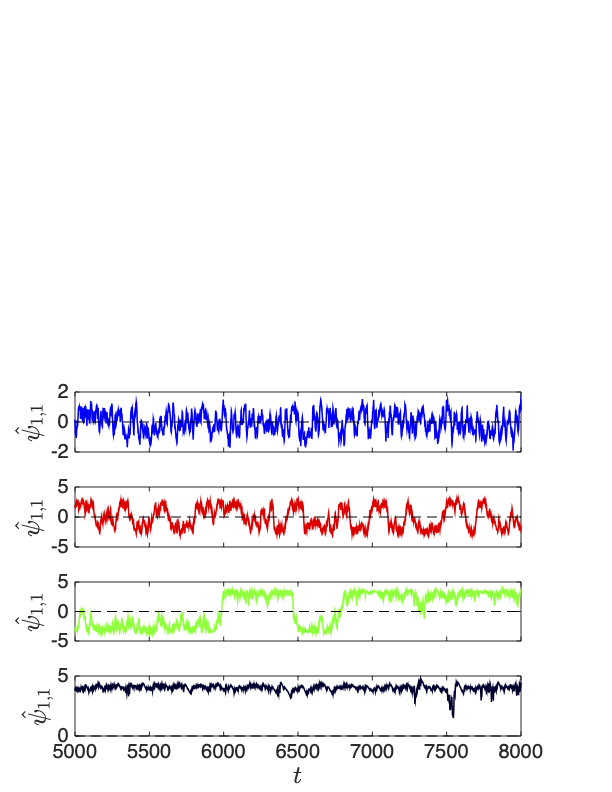}
\includegraphics[width=0.48\textwidth]{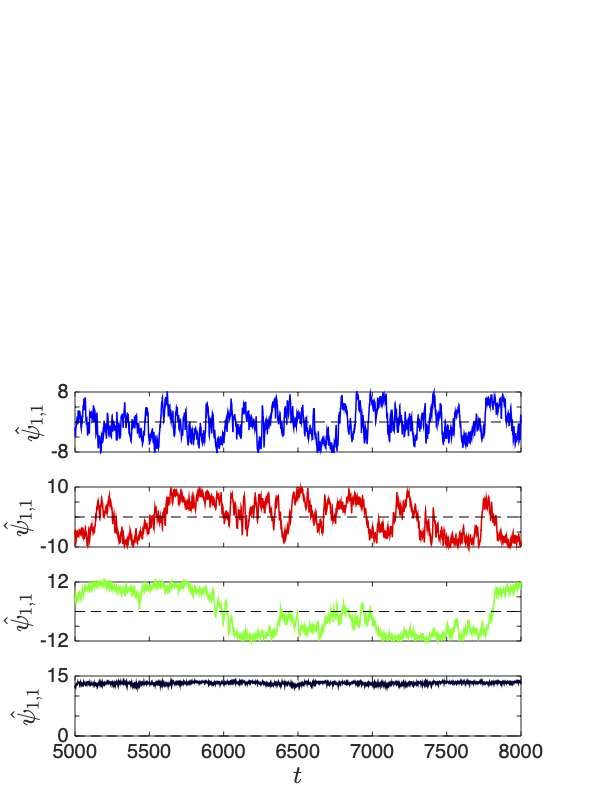}
\caption{(left) Times series of $\hat{\psi}_{1,1}$ obtained from the DNS of the Navier-Stokes equation (NSE) for different values of ${\rm Rh} = 10, 30, 50, 120$ (from top to bottom) and ${\rm Re}=5000/\pi^2$. (right) Times series of $\hat{\psi}_{1,1}$ obtained from the DNS of the truncated-Euler equations (TEE) for different values of $k_c = 5.68, 4.58, 4.10, 3.92$ (from top to bottom). 
$\hat{\psi}_{1,1}^{NSE}$ and $\hat{\psi}_{1,1}^{TEE}$ have been divided by
$10^4$ and $10^2$, respectively. Figure adapted from \cite{shukla2016}, "Copyright (2016) by The American Physical Society".
}
\label{direct recordings}
\end{figure}

Direct numerical simulations of 2D NSE (\ref{eq:NS_incom}) have been performed in Refs. \cite{mishra2015,shukla2016}. The experimental results described above, and in particular the transitions between the different turbulent regimes, have been reproduced. Fig. \ref{direct recordings} (left) displays the direct recording of the large scale mode $\hat{\psi}_{1,1}$ when ${\rm Rh}$ is increased for ${\rm Re} = 5000/\pi^2$. For ${\rm Rh} = 10$, a turbulent flow with a Gaussian large scale velocity is obtained. The flow visualization \cite{mishra2015} shows vorticity concentrations of different sizes that are random in space and time. This corresponds to the regime with an inverse cascade of energy \cite{sommeria1986,paret1997}. When ${\rm Rh} = 30$, the signal fluctuates around two opposite non zero values. The system is in the bimodal regime. When ${\rm Rh} = 50$, the bimodality is more pronounced and reversals between the two opposite polarities of the large scale circulation are clearly visible. The mean waiting time between two successive reversals is much longer than for ${\rm Rh} = 30$. Finally, for ${\rm Rh} = 120$, the system has stoped reversing and stays in one polarity of the large scale circulation.

The large scale drag plays a crucial role in two-dimensional turbulence. First, it is impossible to avoid it in any experimental or natural configuration of quasi 2-D flows. Second, the damping rate $1/\tau$ related to the large scale drag together with the energy flux per unit mass in the inverse cascade, determine the characteristic scale $\lambda_c$ at which the inverse cascade stops.  For large damping rate (${\rm Rh}$ small), $\lambda_c$ is small and the inverse cascade stops before reaching the container size $L$. When the damping rate is small (${\rm Rh}$ large), $\lambda_c$ becomes larger than $L$ and the inverse cascade accumulates kinetic energy at the container scale. The flow is in the condensed state. 

%%%%%%%%%%%%%%%%%%%%%%%
\subsection{Transitions between different turbulent regimes described by the truncated Euler equations}
%%%%%%%%%%%%%%%%%%%%%%%

The two-dimensional truncated Euler equations (TEE) are obtained by setting the right-hand side of Eq.~\eqref{eq:NS_incom} to zero. The TEE conserves two quadratic invariants: the energy,
\begin{equation}
E = \frac{1}{2}\int |\mathbf{u}|^2 \, d^2x,
\end{equation}
and the enstrophy,
\begin{equation}
\Omega = \frac{1}{2} \int |\nabla \times \mathbf{u}|^2 \, d^2x.
\end{equation}
In Fourier space, energy is distributed over modes via the 2D energy spectrum, given by:
\begin{equation}
E(\mathbf{k}) = \frac{1}{2} k^2 |\hat{\psi}(\mathbf{k})|^2,
\end{equation}
where $E(\mathbf{k})$ represents the energy contained in the individual mode $\mathbf{k}$ (without shell averaging).
At late times, the TEE solution reaches a statistically stationary state fully characterized by $E$ and $\Omega$ \cite{kraichnan1967,kraichnan1975}. 

Within the grand canonical framework, the distribution of Fourier amplitudes is assumed to follow:
\begin{equation}
P(\mathbf{u}) = \mathcal{Z}^{-1} \exp(-\alpha E - \beta \Omega),
\label{eq:kraichnan_canonical_pdf}
\end{equation}
where $\alpha$ and $\beta$ are Lagrange multipliers associated with the conservation of energy and enstrophy, respectively, and $\mathcal{Z}$ is a normalization constant.

An alternative is to adopt the microcanonical approach, originally proposed by Lee~\cite{lee1952}, which assumes ergodicity and uniform probability over the phase space constrained by fixed energy $E_0$ and enstrophy $\Omega_0$. The associated probability density function is:
\begin{equation}
P(\mathbf{u}) = \mathcal{Z}^{-1} \delta(E - E_0) \delta(\Omega - \Omega_0),
\label{eq:def_micro_pdf}
\end{equation}
where $\mathcal{Z}$ is again a normalization factor (distinct from that in Eq.~\eqref{eq:kraichnan_canonical_pdf}). Geometrically, this distribution is supported on the intersection of two hypersurfaces-the constant-energy and constant-enstrophy manifolds-within the $N$-dimensional phase space. Because this intersection is a high-dimensional, co-dimension-two manifold, analytical results are generally difficult to obtain from the microcanonical ensemble.

Ref.~\cite{shukla2016} proposed that the TEE can accurately describe the complex transitional dynamics observed in turbulent 2D confined flows subject to bottom friction and periodic forcing. In particular, the large-scale flow reversals seen in such systems-where the large-scale circulation undergoes spontaneous sign changes on top of a turbulent background-can be interpreted as bifurcations in the probability distribution of the large-scale velocity. These transitions, ranging from Gaussian to bimodal distributions and eventual ergodicity breaking, are consistent with the predictions of the microcanonical ensemble. A minimal 13-mode model was shown to reproduce this behavior.

Fig. \ref{direct recordings} (right) displays the time recordings of the large scale mode $\hat{\psi}_{1,1}$ obtained using the TEE and should be compared to the time recordings of the 2D NSE (\ref{eq:NS_incom}) Fig. \ref{direct recordings} (left). 

\begin{figure}
\vspace {-4 cm}
\centering
\includegraphics[width=0.48\textwidth]{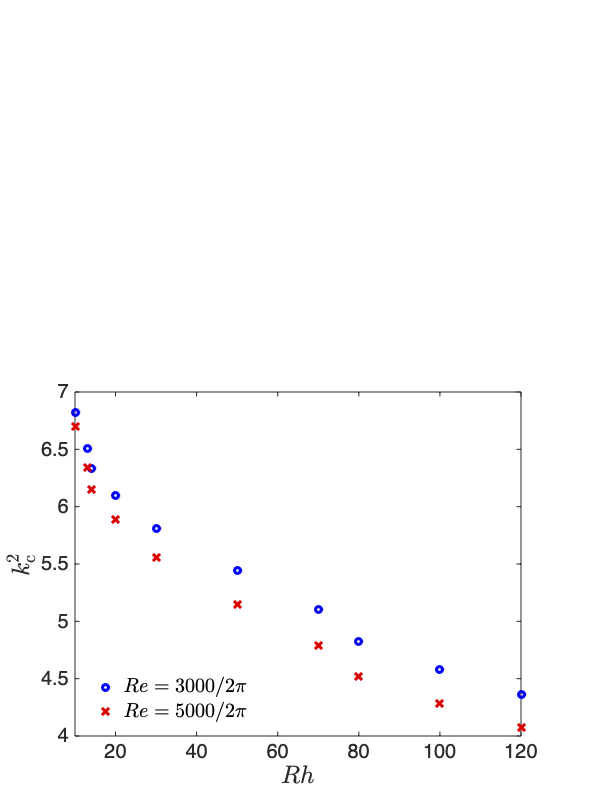}
\includegraphics[width=0.48\textwidth]{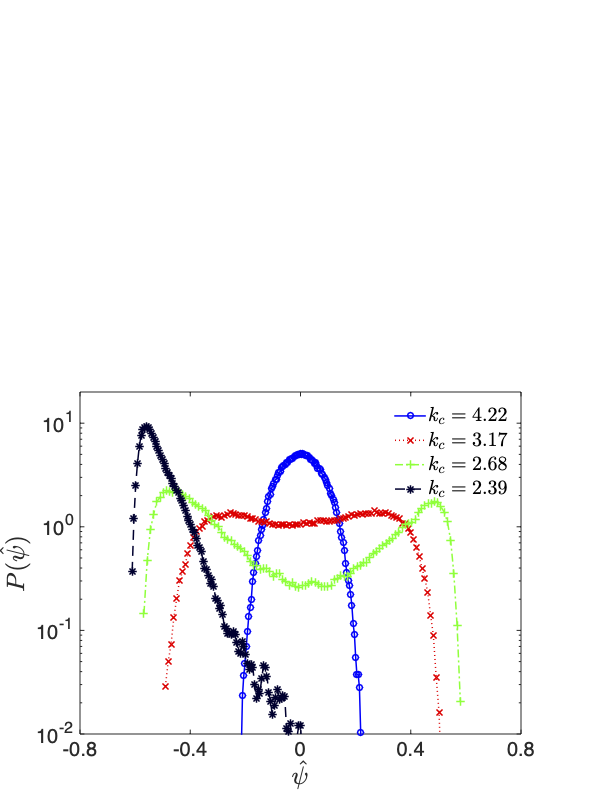}
\caption{(left) Plot of $k_c=\sqrt{\Omega/E}$ versus ${\rm Rh}$ from the DNS of 2D NSE for two different Reynold's numbers ${\rm Re}=3000/\pi^2$ (blue circles) and ${\rm Re}=5000/\pi^2$ (red crosses). (right) Semilogy plots of the PDFs of $\hat{\psi}_{1,1}$ for different values of $k_c$ obtained from the finite-mode minimal model based on TEE. Figure adapted from \cite{shukla2016}, "Copyright (2016) by The American Physical Society".}
\label{k_cRhPDF}
\end{figure}

Of course, the forcing, the viscous dissipation (${\rm Re}$) and the large scale drag (${\rm Rh}$) being absent from TEE, one should identify what is the relevant bifurcation parameter for TEE. The only remaining parameters in the case of TEE, are the two invariants, $E$ and $\Omega$, from which a scale $k_c = (\Omega/E)^{1/2}$ can be constructed. $2\pi/k_c$ plays the role of the scale $\lambda_c$ defined above, at which the inverse cascade stops. This is illustrated in Fig. \ref{k_cRhPDF}(left) which shows that in a direct simulation of  2D NSE, $2\pi/k_c$ increases as ${\rm Rh}$ is increased with a relatviely weak dependence on ${\rm Re}$. Taking into account this correspondance between ${\rm Rh}$ and $k_c$, the TEE reproduce the two transitions between different turbulent regimes observed in experiments and direct simulations. Fig. \ref{k_cRhPDF} (right) shows the probability density functions of $\hat{\psi}_{1,1}$ obtained from the TEE in the different regimes.

Ref.~\cite{vankan2022} advanced the microcanonical theory of two-dimensional TEE by adopting a geometric perspective. The authors explicitly computed phase space volumes over constant-energy and constant-enstrophy shells to evaluate statistical quantities under the microcanonical distribution~\eqref{eq:def_micro_pdf}. This geometric method provides a direct approach to microcanonical statistical mechanics, bypassing the need for a thermodynamic limit.

Two main results were presented. First, the average energy spectrum of highly condensed flows was derived and shown to be consistent with the canonical prediction of Kraichnan, despite the finite dimensionality of the system. Second, extending the analysis of~\cite{shukla2016}, the authors demonstrated that the statistics of flow reversals in the largest-scale mode of a free-slip square-domain flow are well captured by this explicitly geometric microcanonical framework.\\

A approach similar to the one of \cite{shukla2016} has been used to study the bifurcations of large-scale jets in the turbulent regime of a forced two-dimensional shear flow.  It has been also shown that the TEE is able to capture the bifurcations of the large-scale jets exhibited by the Navier-Stokes equations \cite{dallas2020}.\\

\section{Conclusion}
\vskip6pt

We have reviewed in this article several recent works on the dynamical and statistical properties of spatial scales larger than the forcing scale in turbulence. In three-dimensional turbulence, or in analogous wave-turbulence problems involving only a direct cascade of energy, this question has been raised only for about a decade, so experimental and numerical studies remain scarce. All studies concur on the fact that scales larger than the forcing scale are not subject to any non-zero mean energy flux, even when the forcing directly excites large-scale modes through a wave-vector triad involving two forcing modes (spectrally dense forcing).
This means that the large-scale modes, like those of an equilibrium system, receive on average as much energy as they give back. Note however that detailed balance might be broken in the case of spectrally dense forcing.
In any case, the fluctuations of the energy flux are significant. A more detailed investigation of these fluctuations, especially their behaviour as a function of wavenumber, would undoubtedly help clarify the nature of the dominant large-scale mechanisms.
Various studies show that energy equipartition within the large scales is not observed, even approximately, when spectrally dense forcing is used whereas all authors agree on the observation of equipartition when the forcing involves only a few modes at a given wavenumber (spectrally sparse forcing).  The deviation from energy equipartition in the case of spectrally dense forcing results from the direct excitation of large-scale modes through triadic interactions involving two forcing wavevectors. We note that for any given forcing, it is enough to consider a sufficiently large flow domain such that the largest scales cannot be forced in this way. In the limit of a large enough flow volume, we therefore expect energy equipartition to be observed at large scales.

Further study of the dynamical properties of the large scales, following on from the work on decoherence time, for example a comparative study of two-point and two-time correlations for the large scales of turbulence and for truncated Euler equations, would make it possible to assess whether it is reasonable to consider modelling the large scales of certain turbulent flows using truncated Euler equations.\\

Unlike in the case of three-dimensional turbulence, the problem of large scales in two-dimensional turbulence is central and has been the subject of numerous studies, particularly since the discovery of the inverse energy cascade \cite{kraichnan1967}. We have not reviewed all of this literature. At first glance, if we follow the reasoning used for three-dimensional turbulence, it does not seem reasonable to assume that the large scales could be in equilibrium in two-dimensional turbulence, precisely because of the inverse energy cascade. Yet it is in two-dimensional turbulence that equilibrium statistical-physics tools have been used the most extensively. One argument sometimes put forward is that the statistical approach could be applicable in the absence of large-scale drag, which otherwise arrests the inverse cascade before it reaches the scale of the flow domain. In this case, energy accumulates in the largest available mode, and one may imagine that an opposing energy flux becomes significant and cancels out the inverse cascade on average. The large structures thus formed could then be in statistical equilibrium, which would justify the methods used \cite{note_loop}.

A recent study \cite{vankan2025} considered this problem quantitatively by comparing the condensed state obtained in a direct numerical simulation of two-dimensional turbulence without large-scale friction to the predictions of truncated Euler equations. In the case of forcing at constant injected power, it was shown that as the Reynolds number increases, the agreement is good near the threshold for the onset of the condensed state, but a significant disagreement appears at higher Reynolds numbers. It is possible that this result depends on the type of forcing. In the case of Kolmogorov forcing (periodic in space and steady in time), it is known that the injected power can tend to zero as the Reynolds number increases \cite{tsang2009,gallet2013}. The same is obviously true for the dissipated power, and one may wonder whether the system remains close to equilibrium in this case even at high Reynolds number.

In any event, even though we know that the modeling of large scales in two-dimensional turbulence using tools of equilibrium statistical mechanics is a priori invalidated by the inverse cascade, and although we do not expect quantitative agreement on the spectra of the different regimes, except perhaps that of the condensed state, we have presented here a modeling of the large-scale transitions observed between different turbulent regimes of two-dimensional Kolmogorov flows. While it is reasonable to neglect viscous dissipation and forcing when considering the dynamics of scales much larger than the forcing scale, it is nevertheless surprising but remarkable that by accounting for the effect of large-scale friction, by choosing the value of the ratio $\Omega/E$ in the initial conditions of the truncated Euler equations, one observes bifurcations between the different turbulent regimes that agree with those found in experiments and direct numerical simulations.

\ack{This work has been supported by the Agence nationale
de la recherche (Grant No. ANR-23-CE30-0043) and by CNES.}

%%%%%%%%%% Insert bibliography here %%%%%%%%%%%%%%
\bibliography{biblio} 
\bibliographystyle{ieeetr}
%\begin{thebibliography}{9}

%\end{thebibliography}

\end{document}